\documentclass[a4paper]{revtex4}
\usepackage{graphicx}
\usepackage{fancyhdr}
\usepackage{amsmath}
\newcommand{\dis}[1]{\begin{equation}\begin{split}#1\end{split}\end{equation}}
\newcommand{\be}{\begin{equation}}
\newcommand{\ee}{\end{equation}}
\def\bea{\begin{eqnarray}}
\def\eea{\end{eqnarray}}

\newcommand{\eq}[1]{Eq.~(\ref{#1})}

\newcommand{\VEV}[1]{\langle #1 \rangle}

\newcommand{\tev}{\,\textrm{TeV}}
\newcommand{\gev}{\,\textrm{GeV}}

\def\tb{\tan\beta}

\usepackage{color}

\begin{document}

\title{Shifted focus point scenario from 
the minimal mixed mediation of SUSY breaking 
}

%

\author{Bumseok Kyae}
\affiliation{Department of Physics, Pusan National University, Busan 609-735, Korea}

\begin{abstract}
We employ both the minimal gravity- and the minimal gauge mediations of supersymmetry breaking at the grand unified theory (GUT) scale in a single supergravity framework, 
assuming the gaugino masses are generated 
dominantly by the minimal gauge mediation effects \cite{original}.   
In such a ``minimal mixed mediation model,''  
a ``focus point'' of the soft Higgs mass parameter, $m_{h_u}^2$ emerges 
at $3$-$4 \tev$ energy scale, 
which is exactly the stop mass scale needed for explaining the $126 \gev$ Higgs boson mass 
without the ``$A$-term'' at the three loop level. 
As a result, $m_{h_u}^2$ 
can be quite insensitive to various trial stop masses at low energy, 
reducing the fine-tuning measures to be much smaller than $100$ 
even for a $3$-$4 \tev$ low energy stop mass and   
$-0.5 < A_t/m_0\lesssim +0.1$ at the GUT scale. 
The gluino mass is predicted to be about $1.7 \tev$, 
which could readily be tested at LHC run2.
\end{abstract}

\maketitle

\thispagestyle{fancy}


Although the minimal supersymmetric standard model (MSSM) 
has been believed the most promising theory beyond the standard model (SM),  
guiding the SM to a grand unified theory (GUT) or string theory \cite{book,PR1984}, 
any evidence of SUSY has not been observed yet 
at the large hadron collider (LHC).   
The mass bounds on the SUSY particles have gradually increased, and 
now they seem to start threatening the traditional status of SUSY 
as a prominent solution to the naturalness problem of the SM. 
%
Actually, a barometer of the naturalness of the MSSM is the mass of ``stop.''  
Due to the large top quark Yukawa coupling ($y_t$), the top and stop dominantly contribute to the radiative physical Higgs mass squared and also 
the renormalization of a soft mass squared of the Higgs ($m_{h_u}^2$) in the MSSM. 
The renormalization effect on $m_{h_u}^2$ would {\it linearly be sensitive} to  
the stop mass squared, 
while it depends just logarithmically on a ultraviolet (UV) cutoff \cite{book}.  
Since the Higgs mass parameters, $m_{h_u}^2$ and $m_{h_d}^2$ are related to 
the the $Z$ boson mass $m_Z$ 
together with 
the ``Higgsinos'' mass, $\mu$ \cite{book}, 
\dis{ \label{m_Z}
\frac12 m_Z^2=\frac{m_{h_d}^2-m_{h_u}^2{\rm tan}^2\beta}{{\rm tan}^2\beta-1}
-|\mu|^2 , 
}
$\{m_{h_u}^2, m_{h_d}^2, |\mu|^2\}$ should be finely tuned 
to yield $m_Z^2=(91 \gev)^2$ for a given $\tb$ [$\equiv\VEV{h_u}/\VEV{h_d}$], 
if they are excessively large.  
According to the recent analysis based on the three-loop calculations, 
the stop mass required for explaining the $126 \gev$ Higgs boson mass \cite{LHCHiggs} 
without any other helps is about $3$-$4 \tev$ \cite{3-loop}.  
Thus, a fine-tuning of order $10^{-3}$ or smaller looks unavoidable
in the MSSM for a GUT scale cut-off. 

In order to more clearly see the UV dependence of $m_{h_u}^2$ and 
properly discuss this ``little hierarchy problem'', 
however, one should suppose a specific UV model  
and analyze its resulting full renormalization group (RG) equations. 
%
%
One nice idea is the ``focus point (FP) scenario'' \cite{FMM1}. 
It is based on the minimal gravity mediation (mGrM) of SUSY breaking. 
So the soft mass squareds 
such as $m_{h_{u,d}}^2$ and those of the left handed (LH) 
and right handed (RH) stops, ($m_{q_3}^2, m_{u_3^c}^2$) 
as well as the gaugino masses 
$M_a$ ($a=3,2,1$) 
are given to be {\it universal} at the GUT scale, $m_{h_u}^2=m_{h_d}^2=m_{q_3}^2=m_{u_3^c}^2=\cdots\equiv m_0^2$ 
and $M_3=M_2=M_1\equiv m_{1/2}$. 
As pointed out in \cite{FMM1}, 
if the soft SUSY breaking ``$A$-terms'' 
are zero at the GUT scale and 
the unified gaugino mass $m_{1/2}$ 
is just a few hundred GeV, 
$m_{h_u}^2$ converges to a small negative value 
around the $Z$ boson mass scale in this setup, 
{\it regardless of its initial values given by $m_0^2$ at the GUT scale} \cite{FMM1}.
In the RG solution of $m_{h_u}^2$ at the $m_Z$ scale, thus,  
\dis{ \label{RGsol}
m_{h_u}^2(Q=m_Z) = C_{s} m_0^2-C_{g} m_{1/2}^2, 
}
where 
$C_{s}$, $C_{g}$ ($>0$) can numerically be estimated  
using RG equations, 
$C_{s}$ happens to be quite small 
with the above universal soft masses. 
Since stop masses are quite sensitive to $m_0^2$, hence, 
$m_Z^2$ could remain small enough 
even with a relatively heavy stop mass in the FP scenario 
in contrast to the naive expectation. 
  
However, the experimental bound on the gluino mass $M_3$ 
has already exceeded $1.3 \tev$ \cite{gluinomass}. 
As expected from Eqs.~(\ref{m_Z}) and (\ref{RGsol}), 
a too large $m_{1/2}$ needed for $M_3>1.3 \tev$ at low energy 
would require a fine-tuned large $|\mu|$ for $m_Z$ of $91 \gev$ 
particularly for a relatively light stop mass ($\lesssim 1 \tev$) cases. 
When the stop mass is around $3$-$4 \tev$, 
the stop should decouple from the RG equations below $3$-$4 \tev$, 
which makes $C_{s}$ {\it sizable} in \eq{RGsol} \cite{KS}. 
Then, a much larger $m_{1/2}$ is necessary for EW symmetry breaking.  
Since the RG running interval between $3$-$4 \tev$ and $m_Z$ scale, 
to which modified RG equations should be applied, is too large,  
the FP behavior is seriously spoiled    
with such heavy SUSY particles. 

The best way to rescue the FP idea is to somehow shift the FP upto the stop decoupling scale \cite{KS}: 
$C_s$ needs to be made small enough before stops are decoupled.   
Then $m_{h_u}^2$ at the $m_Z$ scale can be estimated using the Coleman-Weinberg potential \cite{book,CQW}.   
It is approximately given by 
\bea \label{RGsm}
m_{h_u}^2(m_Z)
&\approx& m_{h_u}^2(Q_T) - \frac{3|y_t|^2}{16\pi^2}
\left({m}_{q_3}^2+m_{u_3^c}^2\right)\bigg|_{Q_T} , 
\eea
where $Q_T$ denotes the stop decoupling scale. 
Since the $m_0^2$ dependence of stop masses would be loop-suppressed,   
$m_{h_u}^2$ needs to be well-focused around $Q_T$. 
Due to the additional negative contribution to $m_{h_u}^2(m_Z)$ below $Q_T$, 
a small {\it positive} $m_{h_u}^2(Q_T)$ would be more desirable.  
In order to push up the FP to the desired stop mass scale $3$-$4 \tev$, we suggest to combine 
the mGrM and the minimal gauge mediation (mGgM) 
in a single supergravity (SUGRA) framework 
with a {\it common} SUSY breaking source.  
We will call it ``minimal mixed mediation.''

%
%

First, let us consider the minimal K${\rm\ddot{a}}$hler potential, 
and a superpotential where the observable and hidden sectors are separated 
as in the ordinary mGrM \cite{book}:  
\dis{ \label{KSpot} 
K=\sum_{i,a} |z_i|^2+|\phi_a|^2 ~, \quad
W=W_H(z_i)+W_O(\phi_a)
}
where $z_i$ [$\phi_a$] denotes fields in the hidden [observable] sector. 
The kinetic terms of $z_i$ and $\phi_a$, thus, take the canonical form. 
We assume non-zero vacuum expectation values (VEVs) for $z_i$s \cite{PR1984}: 
\dis{ \label{vev}
\langle z_i\rangle=b_iM_P , ~ 
\langle\partial_{z_i} W_H\rangle=a_i^*mM_P , ~ 
\langle W_H\rangle=mM_P^2 ,
}
where $a_i$ and $b_i$ are dimensionless numbers, 
while $M_P$ ($\approx 2.4\times 10^{18} \gev$) is the reduced Planck mass. 
Then, $\langle W_H\rangle$ or $m$ 
gives the gravitino mass, 
$m_{3/2}=e^{K/2M_P}\langle W\rangle/M_P^2=e^{|b_i|^2/2}m$,  
and the ``$F$-terms'' of $z_i$ ($=D_{z_i}W=\partial_{z_i}W+\partial_{z_i}K ~W/M_P^2$)
become of order ${\cal O}(mM_P)$.
The soft terms can read from the scalar potential of SUGRA:
%
when the cosmological constant (C.C.) is fine-tuned to be zero, 
renormalizable terms of it are given by \cite{PR1984} 
\dis{ \label{scalarPot}
V_F \approx \left|\partial_{\phi_a}W_O\right|^2
+m_{0}^2|\phi_a|^2
+m_{0}\left[\phi_a\partial_{\phi_a}W_O
+(A_\Sigma-3)W_O+{\rm h.c.}\right] ,
}
where $A_\Sigma$ is defined as $A_\Sigma\equiv \sum_ib_i^*(a_i+b_i)$, and 
$m_0$ is identified with the gravitino mass $m_{3/2}$ ($=e^{|b_i|^2/2}m$). 
The first term of \eq{scalarPot} is  
the $F$-term potential in global SUSY, the second term is 
the universal soft mass term, and  
the remaining terms are 
$A$-terms,     
which are {\it proportional} to $m_0$.

%

%
    
Next, let us introduce one pair of messenger superfields $\{{\bf 5},\overline{\bf 5}\}$, 
which are the SU(5) fundamental representations. 
Through their coupling with a SUSY breaking source $S$, which is an MSSM  singlet superfield,  
\dis{ \label{Wm}
W_m=y_SS{\bf 5}\overline{\bf 5} ,
}
the soft masses of the MSSM gauginos and scalar superpartners are also radiatively generated \cite{book}:  
\dis{ \label{GGsoft}
M_a= 
\frac{g_a^2}{16\pi^2}\frac{\langle F_S\rangle}{\langle S\rangle} ,~
m_i^2=2 
\sum_{a=1}^{3}
\left[\frac{g_a^2}{16\pi^2}\frac{\langle F_S\rangle}{\langle S\rangle}\right]^2C_a(i) 
}
where $C_a(i)$ is the quadratic Casimir 
invariant for a superfield $i$, $(T^aT^a)_i^j=C_a(i)\delta_i^j$, and   
$g_a$ ($a=3,2,1$) denotes the MSSM gauge couplings. 
$\langle S\rangle$ and $\langle F_S\rangle$ are VEVs of the scalar and $F$-term components of the superfield $S$. 
The mGgM effects would appear below the messenger scale, $y_S\langle S\rangle$.  
Here we assume that $\langle S\rangle$ has the same magnitude as
the VEV of the SU(5) breaking Higgs $v_G$: 
$\langle {\bf 24}_H\rangle=v_G\times {\rm diag.}(2,2,2;-3,-3)/\sqrt{60}$. 
It is possible if a GUT breaking mechanism causes $\langle S\rangle$. 
Actually, the ``$X$'' and ``$Y$'' gauge boson masses,  
$M_X^2=M_Y^2=\frac{5}{24}g_G^2v_G^2$ \cite{GUT}, 
where $g_G$ is the unified gauge coupling,  
can be identified with the MSSM gauge coupling unification scale. 

%
%

\begin{figure}[ht]
\centering
\includegraphics[width=80mm]{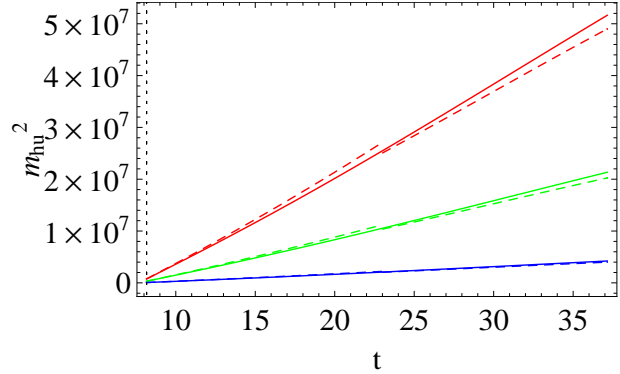}
\caption{RG evolutions of $m_{h_u}^2$ 
with $t$ [$\equiv {\rm log}(Q/\gev)]$
for $m_0^2=(7 \tev)^2$ [Red], $(4.5 \tev)^2$ [Green], 
and $(2 \tev)^2$ [Blue] when $A_t=-0.2 ~m_0$ and $\tb=50$. 
The tilted straight [dotted] lines correspond to the case of 
$t_M\approx 37$ (or $Q_M\approx 1.3\times 10^{16} \gev$, ``{\rm Case A}'') 
[$t_M\approx 23$ (or $Q_M=1.0\times 10^{10} \gev$, ``{\rm Case B}'')].
The vertical dotted line at $t=t_T\approx 8.2$ ($Q_T=3.5 \tev$) indicates 
the desired stop decoupling scale. 
The discontinuities of $m_{h_u}^2(t)$ should appear at the messenger scales.
As seen in the figure, the FP scale is not affected by messenger scales. 
%
%
} \label{fig:1}
\end{figure}

In addition to \eq{KSpot}, the K${\rm\ddot{a}}$hler potential 
(and hidden local symmetries we don't specify here) 
can permit     
\dis{ \label{Extpot}
K\supset f(z)S +{\rm h.c.} ,
}      
where $f(z)$ denotes a {\it holomorphic} monomial of hidden sector fields $z_i$s 
with VEVs of order $M_P$ in \eq{vev}, 
and so it is of order ${\cal O}(M_P)$. 
Their kinetic terms still remain canonical.  
The U(1)$_R$ symmetry forbids $M_Pf(z)S$ in the superpotential.
Then, the resulting $\langle F_S\rangle$ can be  
$\langle F_S\rangle\approx m\left[\langle f(z)\rangle+\langle S^*\rangle\right]$
by including the SUGRA corrections with $\langle W_H\rangle=mM_P^2$. 
Thus, the VEV of $F_S$ is of order ${\cal O}(mM_P)$ like $F_{z_i}$. 
They should be fine-tuned for the vanishing C.C.:  
a precise determination of $\langle F_S\rangle$ is indeed 
associated with the C.C. problem.
%
%
Here we set $\langle F_S\rangle=m_0M_P$. 
%
%
%
Thus, the typical size of mGgM effects is estimated as
\dis{ \label{GM}
\frac{\langle F_S\rangle}{16\pi^2\langle S\rangle}=\frac{m_0 M_P}{16\pi^2M_X}\sqrt{\frac{5}{24}}g_{G}
\approx 0.36 \times m_0 ,  ~~~ 
}
where $g_G$ is set to be $\sqrt{4\pi/26}$ at the GUT scale 
[$\approx (1.3\pm 0.4)\times 10^{16} \gev$] 
due to relatively heavy colored superpartners ($\gtrsim 3 \tev$). 
Even for $|y_S|\ll 1$, we will keep this value,   
since it is fixed by a UV model.
For $|y_S|\lesssim 1$ in \eq{Wm}, 
the messenger scale $Q_M$ drops down below $M_{X,Y}$.  
The soft masses generated by the mGgM in \eq{GGsoft} 
are {\it non-universal} for $Q_M<M_{X,Y}$, and the beta function coefficients of the MSSM fields should be modified 
above the $Q_M$ scale by  
the messenger fields $\{{\bf 5}, \overline{\bf 5}\}$. 
%
%
%
The boundary conditions at the GUT scale are of the universal form 
as seen in \eq{scalarPot}. 
We have {\it additional} non-universal contributions 
by \eq{GGsoft}. 
They should be imposed at a given messenger scale,  
and so affect the RG evolutions of MSSM parameters for $Q\leq Q_M$.
%
%

We also suppose that the gaugino masses from the mGrM  
are relatively suppressed. 
In fact, the gaugino mass term in SUGRA is associated with the first derivative of the gauge kinetic function \cite{PR1984}, 
and so a constant gauge kinetic function at tree level 
($= \delta_{ab}$) can realize it.  
Thus, the gaugino masses by \eq{GGsoft} dominates over them in this case.  
Then, a simple analytic expression 
for the gaugino masses at the stop mass scale is possible:  
$M_a(Q_T)\approx 0.36\times m_0\times g_a^2(Q_T)$.
It does {\it not} depend on messenger scales, $A_t$, $\tb$, etc.  

The fact that the mGgM effects by \eq{GGsoft} are proportional to $m_0$ or $m_0^2$
are important. 
Moreover, $A$-terms from \eq{scalarPot} are also proportional to $m_0$.  
In this setup, thus,    
an (extrapolated) FP of $m_{h_u}^2$ must still exist
at a higher energy scale.    
As $C_g$ is converted to a member of $C_s$ in \eq{RGsol}, 
the naturalness of $m_{h_u}^2$ and $m_Z^2$ becomes gradually improved, 
making $C_s$ smaller and smaller, 
until the FP reaches the stop decoupling scale.

%
Fig.~\ref{fig:1} displays RG evolutions of $m_{h_u}^2$ under various trial $m_0^2$s. 
%
The straight [dotted] lines correspond to the case of 
$t_M\approx 37$ (or $Q_M\approx 1.3\times 10^{16} \gev$, ``Case A'') 
[$t_M\approx 23$ (or $Q_M=1.0\times 10^{10} \gev$, ``Case B'' )].
The discontinuities of the lines by additional boundary conditions arise at the messenger scales. 
%
As seen in Fig.~\ref{fig:1}, a FP of $m_{h_u}^2$ appears always at $t=t_T\approx 8.2$ (or $Q_T\approx 3.5 \tev$) {\it regardless of the chosen messenger scales}. 
%
%
Hence, the wide ranges of UV parameters can yield almost the same values of $m_{h_u}^2$ at low energy.  
Under this situation, one can guess that $m_0^2\approx (4.5 \tev)^2$ happens to be selected, yielding $3$-$4 \tev$ stop mass, 
and so eventually gets responsible for the $126 \gev$ Higgs mass.  
In both cases of Fig.~\ref{fig:1}, 
the low energy gaugino masses are
\dis{ \label{lowM_a}
M_{3,2,1}\approx \{1.7 \tev, ~660 \gev, ~360 \gev\} 
}
for $m_0^2=(4.5 \tev)^2$. 
They would be testable at LHC run2.
$A_t$ at low energy is about 
$1 \tev$ for Case A and B. 
So the contributions of $A_t^2/\widetilde{m}_t^2$ to the radiative Higgs mass 
are smaller than 
2.3 $\%$ of 
those by the stops.

%
%
%
%

%
\begin{table}[ht]
\begin{center}
\caption{Soft squared masses of the stops and Higgs bosons at $t=t_T\approx 8.2$ ($Q_T=3.5 \tev$) for various trial $m_0^2$s 
when $Q_M\approx 1.3\times 10^{16} \gev$. 
$\Delta_{m_0^2}$ indicates the fine-tuning measure for $m_0^2=(4.5 \tev)^2$ 
in each case. 
$m_{h_u}^2$s further decrease to be negative below $t=t_T$.    
The mass spectra are generated using SOFTSUSY  \cite{softsusy}.
}
\begin{tabular}
{c|ccc||c|ccc}
\hline\hline
 {\bf Case I} & {\footnotesize $A_t=0$}  & {\footnotesize $\tb=50$}~  & {\footnotesize ${\bf \Delta_{m_0^2}=1}$}  
&
 {\bf Case II} & {\footnotesize $A_t=-0.2~m_0$}  & {\footnotesize $\tb=50$}~  & {\footnotesize ${\bf \Delta_{m_0^2}=16}$}  
\\ 
 {\footnotesize ${\bf m_0^2}$} & {\footnotesize $({\bf 5.5} \tev)^2$}  & {\footnotesize $({\bf 4.5} \tev)^2$}  & {\footnotesize $({\bf 3.5} \tev)^2$}  
&
 {\footnotesize ${\bf m_0^2}$} & {\footnotesize $({\bf 5.5} \tev)^2$}  & {\footnotesize $({\bf 4.5} \tev)^2$}  & {\footnotesize $({\bf 3.5} \tev)^2$}  
\\ \hline
 {\footnotesize $m_{q_3}^2(t_T)$} &  {\footnotesize $(4363 \gev)^2$}  & {\footnotesize $(3551 \gev)^2$}  & {\footnotesize $(2744 \gev)^2$} 
&
 {\footnotesize $m_{q_3}^2(t_T)$} &  {\footnotesize $(4376 \gev)^2$}  & {\footnotesize $(3563 \gev)^2$}  & {\footnotesize $(2752 \gev)^2$} 
\\
 {\footnotesize $m_{u^c_3}^2(t_T)$} &  {\footnotesize $(3789 \gev)^2$}  & {\footnotesize $(3098 \gev)^2$}  & {\footnotesize $(2406 \gev)^2$} 
&
 {\footnotesize $m_{u^c_3}^2(t_T)$} &  {\footnotesize $(3798 \gev)^2$}  & {\footnotesize $(3106 \gev)^2$}  & {\footnotesize $(2413 \gev)^2$} 
\\
 {\footnotesize ${\bf m_{h_u}^2(t_T)}$} & {\footnotesize $({\bf 431} \gev)^2$}  & {\footnotesize $({\bf 189} \gev)^2$}  & {\footnotesize $-({\bf 251} \gev)^2$}  
&
 {\footnotesize ${\bf m_{h_u}^2(t_T)}$} & {\footnotesize $({\bf 539} \gev)^2$}  & {\footnotesize $({\bf 361} \gev)^2$}  & {\footnotesize $-({\bf 44} \gev)^2$}  
\\
 {\footnotesize $m_{h_d}^2(t_T)$} &  {\footnotesize $(2022 \gev)^2$}  & {\footnotesize $(1512 \gev)^2$}  & {\footnotesize $(1008 \gev)^2$}
&
 {\footnotesize $m_{h_d}^2(t_T)$} &  {\footnotesize $(2053 \gev)^2$}  & {\footnotesize $(1565 \gev)^2$}  & {\footnotesize $(1046 \gev)^2$}
\\ \hline\hline
 {\bf Case III} & {\footnotesize $A_t=-0.5~m_0$}  & {\footnotesize $\tb=50$}~  & {\footnotesize ${\bf \Delta_{m_0^2}=9}$}  
&
 {\bf Case IV} & {\footnotesize $A_t=0$}  & {\footnotesize $\tb=25$}~  & {\footnotesize ${\bf \Delta_{m_0^2}=57}$}  
\\ 
 {\footnotesize ${\bf m_0^2}$} & {\footnotesize $({\bf 5.5} \tev)^2$}  & {\footnotesize $({\bf 4.5} \tev)^2$}  & {\footnotesize $({\bf 3.5} \tev)^2$}  
&
 {\footnotesize ${\bf m_0^2}$} & {\footnotesize $({\bf 5.5} \tev)^2$}  & {\footnotesize $({\bf 4.5} \tev)^2$}  & {\footnotesize $({\bf 3.5} \tev)^2$}  
\\ \hline
 {\footnotesize $m_{q_3}^2(t_T)$} &  {\footnotesize $(4284 \gev)^2$}  & {\footnotesize $(3532 \gev)^2$}  & {\footnotesize $(2630 \gev)^2$} 
&
 {\footnotesize $m_{q_3}^2(t_T)$} &  {\footnotesize $(4915 \gev)^2$}  & {\footnotesize $(4025 \gev)^2$}  & {\footnotesize $(3134 \gev)^2$} 
\\
 {\footnotesize $m_{u^c_3}^2(t_T)$} &  {\footnotesize $(3755 \gev)^2$}  & {\footnotesize $(3088 \gev)^2$}  & {\footnotesize $(2373 \gev)^2$} 
&
 {\footnotesize $m_{u^c_3}^2(t_T)$} &  {\footnotesize $(3770 \gev)^2$}  & {\footnotesize $(3086 \gev)^2$}  & {\footnotesize $(2400 \gev)^2$} 
\\
 {\footnotesize ${\bf m_{h_u}^2(t_T)}$} & {\footnotesize $-({\bf 363} \gev)^2$}  & {\footnotesize $-({\bf 41} \gev)^2$}  & {\footnotesize $-({\bf 546} \gev)^2$}  
&
 {\footnotesize ${\bf m_{h_u}^2(t_T)}$} & {\footnotesize $({\bf 152} \gev)^2$}  & {\footnotesize $-({\bf 220} \gev)^2$}  & {\footnotesize $-({\bf 293} \gev)^2$}  
\\
 {\footnotesize $m_{h_d}^2(t_T)$} &  {\footnotesize $(1447 \gev)^2$}  & {\footnotesize $(1359 \gev)^2$}  & {\footnotesize $-(950 \gev)^2$}
&
 {\footnotesize $m_{h_d}^2(t_T)$} &  {\footnotesize $(5057 \gev)^2$}  & {\footnotesize $(4136 \gev)^2$}  & {\footnotesize $(3215 \gev)^2$}
\\ \hline\hline
\end{tabular}
\label{tab:data}
\end{center}
\end{table}
%

Table~\ref{tab:data} lists the soft squared masses at $t=t_T$ 
for the LH and RH stops, and the two MSSM Higgs bosons under the various $m_0^2$s, 
when $Q_M\approx 1.3\times 10^{16} \gev$, and $\tb$ is $50$ or $25$. 
We can see the changes of $m_{h_u^2}^2$ are quite small [$\ll (550 \gev)^2$] under the changes of $m_0^2$ [$(5.5 \tev)^2$--$(3.5 \tev)^2$], 
because $m_{h_u}^2$ is well-focused at $t=t_T$.   
Case I-IV yield again the same low energy gauginos masses as \eq{lowM_a}. 
$A_t$ at low energy turns out to be around $1 \tev$ or smaller 
for $m_0^2=(4.5 \tev)^2$, 
and so its contribution to the Higgs boson mass is still suppressed.
By \eq{RGsm} $m_{h_u}^2$s further decrease to be negative below $t=t_T$.     
With \eq{m_Z} $|\mu|$ are determined as 
$\{485 \gev, 392 \gev, 516 \gev, 586 \gev\}$ 
for Case I, II, III, and IV, respectively.
%
%
%
%
In Table~\ref{tab:data}, 
the fine-tuning measure $\Delta_{m_0^2}$ 
($\equiv \left|\frac{\partial{\rm log}m_Z^2}{\partial {\rm log}m_0^2}\right|
=\left|\frac{m_0^2}{m_Z^2}\frac{\partial m_Z^2}{\partial m_0^2}\right|$ \cite{FTmeasure}) 
around $m_0^2=(4.5 \tev)^2$ are also presented. 
They are of order ${\cal O}(1-10)$. 
$\Delta_{A_t}$ ($=\left|\frac{A_t}{m_Z^2}\frac{\partial m_Z^2}{\partial A_t}\right|$) is estimated as $\{0, 10, 118, 0\}$ 
for Case I, II, III, and IV, respectively. 
When $A_t/m_0=+0.1$, $\{\Delta_{m_0^2}, \Delta_{A_t}, |\mu|\}$ turn out to be about $\{22, 33, 569 \gev\}$. 
Therefore, the parameter range 
\dis{
-0.5 ~<~  A_t/m_0 ~\lesssim~ +0.1 
~~~{\rm and}~~ \tb \gtrsim 25  
}  
allows $\{\Delta_{m_0^2}, \Delta_{A_t}\}$ and $|\mu|$ 
to be smaller than $100$ 
and $600 \gev$, respectively \cite{original}. 
We see that a larger $\tb$ would be preferred for a smaller $\Delta_{m_0^2}$. 
It is basically because $m_{h_d}^2$ is not focused unlike $m_{h_u}^2$, 
even if it also contributes to $m_Z^2$ as seen in \eq{m_Z}. 
$\tb=50$ is easily obtained e.g. from the minimal SO(10) GUT \cite{GUT}.  
%
%
%
%

In conclusion, we have noticed that a FP of $m_{h_u}^2$ appears at $3$-$4 \tev$, 
when the mGrM and mGgM effects are combined at the GUT scale 
for a common SUSY breaking source parametrized with $m_0$, 
and the gaugino masses are dominantly generated by the mGgM effects. 
Even for a $3$-$4 \tev$ stop mass explaining the $126 \gev$ Higgs mass, 
thus, the fine-tuning measures significantly decrease well below $100$ 
for $-0.5< A_t/m_0\lesssim +0.1$ and $\tb \gtrsim 25$ 
in the minimal mixed mediation.   
In this range, $|\mu|$ is smaller than $600 \gev$.     
The expected gluino mass is about $1.7 \tev$, 
which could readily be tested at LHC run2.

%


\bigskip 


\end{document}